\documentstyle[12pt,twoside]{article}
\def\greaterthansquiggle{\raise.3ex\hbox{$>$\kern-.75em\lower1ex\hbox{$\sim$}}}
\def\lessthansquiggle{\raise.3ex\hbox{$<$\kern-.75em\lower1ex\hbox{$\sim$}}}
\newcommand{\beq}{\begin{equation}}
\newcommand{\eeq}{\end{equation}}
\newcommand{\beqa}{\begin{eqnarray}}
\newcommand{\eeqa}{\end{eqnarray}}
\newcommand{\beqan}{\begin{eqnarray*}}
\newcommand{\eeqan}{\end{eqnarray*}}
\newcommand{\ba}{\begin{array}}
\newcommand{\ea}{\end{array}}

\newcommand{\A}{{\cal A}}
\newcommand{\B}{{\cal B}}

\newcommand{\M}{{\cal M}}

\def\nz{\ifmmode {I\hskip -3pt N} \else {\hbox {$I\hskip -3pt N$}}\fi}
\def\zz{\ifmmode {Z\hskip -4.8pt Z} \else
       {\hbox {$Z\hskip -4.8pt Z$}}\fi}
\def\qz{\ifmmode {Q\hskip -5.0pt\vrule height6.0pt depth 0pt
       \hskip 6pt} \else {\hbox
       {$Q\hskip -5.0pt\vrule height6.0pt depth 0pt\hskip 6pt$}}\fi}
\def\rz{\ifmmode {I\hskip -3pt R} \else {\hbox {$I\hskip -3pt R$}}\fi}
\def\cz{\ifmmode {C\hskip -4.8pt\vrule height5.8pt\hskip 6.3pt} \else
       {\hbox {$C\hskip -4.8pt\vrule height5.8pt\hskip 6.3pt$}}\fi}

\newtheorem{definition}{Definition}

\def\au{{\setbox0=\hbox{\lower1.36775ex%
\hbox{''}\kern-.05em}\dp0=.36775ex\hskip0pt\box0}}
\def\ao{{}\kern-.10em\hbox{``}}

\normalsize
\begin{document}
\bibliographystyle{plain}

\begin{titlepage}
\begin{flushright}
\today
\end{flushright}
\vspace*{2.2cm}
\begin{center}
{\Large \bf D-grading and quasifree evolution }\\[30pt]

Heide Narnhofer  $^\ast $\\ [10pt] {\small\it}
Fakult\"at f\"ur Physik \\ Universit\"at Wien\\

\vfill \vspace{0.4cm}

\begin{abstract}Generalizing the relation between spin-systems and Fermi-systems on the lattice we construct for a spin-system with dimension d an algebra for which  quasifree time-evolutions exist. With appropriate assumptions the gauge invariant subalgebra common for both algebras is invariant under this time evolution and on this subalgebra is norm asymptotically abelian.

\smallskip
Keywords:  d-grading, commutation relations, asymptotic behaviour
\\
\hspace{1.9cm}

\end{abstract}
\end{center}

\vfill {\footnotesize}

$^\ast$ {E--mail address: heide.narnhofer@
univie.ac.at}
\end{titlepage}
\section{Introduction}
Already in \cite{A}, \cite{AMo} 2-grading, i.e distinguishing between even and odd operators that is defined for quantum spin systems with spin-dimension 2 and for the Fermi- algebra built by creation and annihilation operators allows a passage between these two algebras that is able to transfer time evolution from one algebra to the other. Especially gauge invariant time automorphisms on the Fermi-algebra exist, since they are related to time evolution on the lattice defined with using convergence properties \cite{AMo}. On the other hand for the XY-model the relation of the time evolution to a quasifree evolution of the Fermi-algebra allows to control asymptotics as well as symmetry breaking \cite{AMa} and this relation still offers new results \cite{O} or recently \cite{RMO}. This connection can be generalized with some restrictions to other quasifree evolutions

It is natural to ask how far these considerations can be transferred to d-grading. Especially we are interested to obtain results for the asymptotic behaviour of time evolutions for the spin lattice system with spin-dimension $d.$ In \cite{N1},\cite{N2} it was shown that for time evolution based on local hamiltonians as constructed in \cite{BR}  an operator can be found that does not commute asymptotically in time with all other operators. Therefore the time evolution is not norm asymptotically abelian for the total algebra. But the special time evolution given in the XY-model is norm asymptotically abelian on the even subalgebra. We are interested to find for spins of higher dimension such time evolutions that are asymptotically abelian for the corresponding subalgebra. Also we are interested whether the possibility to consider spins of dimension $2$ as being coarse grained from spins of higher dimension offers the possibility to construct time evolutions that are asymptotically abelian on a smaller algebra.

In this paper we follow a variation of the strategy in \cite{A}, \cite{AMa}: There the Fermi-algebra was extended by the crossed product of the grading automorphism. In this algebra the Fermi-algebra can be imbedded, but also the spin-algebra. The intersection of the two algebras is the even subalgebra of both algebras. We start instead from the beginning with the subalgebra of the spin system invariant under the automorphism corresponding to rotation with respect to one spin direction and define it as the gauge invariant subalgebra. This algebra is enlarged by the crossed product with an appropriately chosen automorphism group given by $\gamma $ satisfying $\gamma^d=1$. With the appropriate choice of $\gamma $ the crossed product becomes the total spin algebra. But other choices are possible, and we find the one $\M_F (d)$ that corresponds to the Fermi-algebra, i.e. space translations on this algebra are well defined and the algebra is built by operators that have a particle-structure with non-trivial commutation relations, namely those that generate the automorphism $\gamma.$ We can control the commutation relations between the space translated generators of $\gamma $ in a way that defines a linear relation among them. On this linear space  continuous evolutions can be defined that commute with space translation. We express at least formal unitaries that implement this automorphism explicitely  and observe the relevance with reflection as antiisomorphism. This explicit construction is a generalization to the familiar one for $2-$grading to $d-$ grading  and allows to define a continuous space translation for the $\M_F (d)$ algebra. As for $d=2$ \cite{MN} this continuous automorphism cannot be transferred to the spin algebra on the lattice. But continuous automorphisms in analogy to quasifree automorphisms can be constructed that with some restrictions act as automorphisms on the gauge invariant subalgebra and  can be extended to the spin algebra.   We formulate the restrictions of these evolutions . In addition the asymptotics of these evolutions can be controlled and lead to a d-commutativity on the extended  $\M_F$-algebra, asymptotic abelianess on the gauge invariant algebra but is definitely not asymptotically abelian for the extended spinalgebra. In addition time translation invariant states apart from KMS-states can be given. Also cluster properties and restrictions on expectation values follow.

In section1 we construct the various d-graded algebras based on Weyl-operators and possible automorphims $\gamma $ created by these Weyl-operators as crossed product of a common gauge invariant subalgebra with these automorphisms. In section 2 we control the commutation relations of the operators implementing the automorphism
$\gamma $ respectively its shifted version. This allows to define the algebra via operators over a linear space replacing the creation and annihilation operators of the Fermi-algebra. In section 3 we define the continuous extension of the shift for the Fermi-algebra together we the fact that it is not defined on the spin-algebra. In section 4 we study the  conditions on unitary operators on this linear space so that they correspond to an automorphism on the algebra. Especially we are interested in the asymptotic behaviour of these automorphisms. In section 5  we study the relation between  Fermi-algebras of $d$-grading and $kd$-grading and show their equivalence. We use the possibility to transfer automorphisms from one algebra to the other inheriting control on the asymptotic behaviour as in \cite{AMa}. As a consequence we find new continuous automorphisms on the spin algebra that are not norm asymptotic abelian on the total algebra but with increasing generalisation on the permitted automorphisms are norm asymptotically abelian on a decreasing set of subalgebras. However this is only possible by violating the invariance of the time evolution under space translation with steps corresponding to $d-$grading. Finally in section 6 we offer how the construction can be generalized to algebras on a higher dimensional lattice by replacing the calculations on the formal unitaries by using the antiisomorphism of reflection and the crossed product construction for antiisomorphisms.

\section{The gauge invariant quasilocal algebra and its extension as crossed product}

We start with a matrix-algebra $M_0$ of dimension d  and its tensor product $\M=\otimes _{x\in Z} M_x $ with the space translated $M_x =\sigma _x M_0$ as the $C^*$ algebra as norm closure  of the corresponding algebra with $x\in \Lambda$, $\Lambda $ finite subsets of $Z.$ $M_0$ is created by Weyl-operators $W(k,l)$ satisfying
 \beq W(k,l)W(m,n)=e^{\pi i \frac{kn-lm}{d}}W(k+m,l+n), \quad W(d,0)=1,\eeq $$W(1,0)W(0,1)W(0,-1)=e^{2\pi i/d}W(0,1)$$
As a subalgebra of $\M$ we take the algebra $\A$ which is invariant under all automorphisms implemented by $W_x(1,0) \quad \forall x \in Z$. It consists of products of $W_x(1,0)$ together $W_x(0,1)W_y(0,-1)$ that can be located at different points and products of these operators. This subalgebra has a quasilocal structure, i. e.
$ \A_{[r,s]}$ commutes with $\A_{[r+n,s+n]}, r+n > s.$
Then ${\M}$ is the crossed product with the automorphism $\gamma $ implemented by $W_0(0,1)$. We add simply $W_0(0,1)$ when creating the operators in $M$, referring to the crossed product construction only has the advantage that we control the topology as a C*-algebra. Here the crossed product is given as the algebra ${\B}_{ \gamma }$ consisting of  operators $(b_1,b_2)$ with multiplication rule
\beq (b_1,b_2)(c_1,c_2)=(b_1c_1+b_2\gamma c_2, b_1c_2+b_2\gamma c_1)\eeq
so that $\A_{\gamma }=\M.$

However other choices  $\gamma _{\beta } $ are possible. $\gamma_ {\beta } $ has to be a non-inner automorphism on $\A$. To preserve grading we demand $\gamma_{\beta }^d=1$. To preserve the local structure we demand that $\sigma _x\gamma _{\beta } \sigma _{-x} \gamma _{\beta }^{-1}$ has to be an inner automorphism of $\A.$ Evidently $\gamma $ satisfies all these demands. Another possibility is offered by

\begin{definition}$\gamma _{\beta }= \gamma  \beta _+  \beta _-$
with $\beta _+ $ acting on $\A_{[1,\infty ]}$ as rotation with $W(1,0)^{j_+}$ and $\beta _-$ acting on $ \A _{[-1, -\infty ]}$ as rotation with $W(1,0)^{j-}$ at every point.\end{definition}
$ \gamma$ satisfies all demands and commutes with $\beta _+ \beta _-$ . Further combined with spacetranslation $\sigma _x$
$$ \sigma _x \beta _+\beta _-\sigma _{-x} \beta _+^{-1}\beta _-^{-1}$$
is formally implemented by

\beq\Pi _{0< y < x} W_y(1,0)^{(j_- -j_+)}W_0(1,0)^{j_-}W_x(1,0)^{j_+}\eeq
which acts therefore only locally and on $\A_{[\Lambda]}$ is an inner automorphism. It satisfies grading if $j_+ -j_-= 0,1,2,..d-1.$ Notice, that for $j_+  j_-\neq 0$ $\beta _+\beta _-$ is not an inner automorphism for $\M_.$ Therefore we need the formalism of the crossed product to be sure that the algebra is well defined. Nevertheless we can think of $\gamma _{\beta }$ to be implemented by
\beq \bar{W}(0,1) =W_0(0,1) \otimes \Pi _{x>0}W_x(1,0)^{j_+}\otimes\Pi _{y<0}W_y(1,0)^{j_-}\eeq
where the operators $W_x$ are Weyl-operators in $\sigma _x M_0.$ This allows to define formally $\bar{W}_x$ as the space translated operator that implements the automorphism $\sigma _x \gamma \beta \sigma _{-x}.$  Combined
\beq \bar{W}_x(0,1)\bar{W}_y(0,1)=\eeq $$W_x(0,1)W_x(1,0)^{j_+}W_y(1,0)^{j_-}W_y(0,1)\Pi _{0<z<x}W_z(1,0)^{j_x+j_+} \Pi _{z<0}W_z(1,0)^{2j_-}\Pi _{z>x}W_z(1,0)^{2j_-} $$
Though this is not a well defined operator in $\M$ it defines an automorphism on $\M$ and on its restriction to $\A$ and allows the construction of $\M_{\beta }$ as the crossed product with the automorphism $\gamma _{\beta }.$
Further we obtain the commutation rule
$$\bar{W}_x(0,1)\bar{W}_y(0,1)=e^{2\pi i\/d}\bar{W}_y(0,1)\bar{W}_x(0,1) \quad x<y$$
The explicit connection between the gauge invariant subalgebras is given by
$$\bar{W}_x(0,1)\bar{W}_y(0,-1)=W_x(0,1)W_x(1,0)^{j_+}\Pi _{x<z<y} W_z(1,0)^{j_++j_-}W_y(1,0)^{-j_+}W_y(0,-1) $$
Therefore the local algebras $\A_{[r,s]}$ coincide though with a mapping that does not correspond to an automorphism.
In the same way as $W(r,s )$ can be constructed to obtain matrix units we obtain matrix units in the crossed product. With $\bar{W}(r,s)=W(r,0)\bar{W}(0,s)$ the same linear combinations as for $W(r,s)$ give the new matrix units. They can be represented in the same Hilbertspace corresponding to the groundstate where the groundstate is implemented by $|1,1..>$ and other vectors are given by $|s_1,...s_x,..>, s_n=1 \forall n>n_0$ and their limits with respect to the scalarproduct $<s_1,  s_x,..|t_1,  t_x>=\Pi_j \delta _{s_j,t _j}.$ The representation is given by

\beq m_x(r,s)|t_1,..t_x,...>=\delta _{s,t_x}m_x(r,s)|t_1,..s,..>= \delta _{s,t_x}|t_1,...r,..> \eeq
respectively by

\beq \bar{m}_x(r,s)|t_1,..t_x,...>=\delta _{s,t_x}\bar{m}_x(r,s)|t_1,...t_{x-1},s,..>=\delta _{s,t_x}c(t_1,..t_{x-1})|t_1,..r,..> \eeq
with $c$ the appropriate phase factor given by the commutation rules. Therefore $\bar{m}_0(r,s)$  can be considered as replacements of creation and annihilation operators for $d=2.$ However the exact value of $c$ is less transparent. It is based on the commutation rule for the Weyl-operators taking into account the additional action of $\sigma^{\pm i/d}$, namely
\beq \bar{W}_x(0,1)\bar{W}_y(0,1)-e^{2\pi i/d}\bar{W}_y \bar{W}_x(0,1)=0,\quad x<y\eeq

 Calculating the passage to the commutation rules for the matrix units we obtain:
\beq \bar{m}_x(j,k)\bar{m}_y(l,n)=e^{i\pi (j-k-l+n)(x-y)}\bar{m}_y(l,n)\bar{m}_x(j,k),\quad x\neq y\eeq
The norm of the operators $\bar{m}_x(k,l)$ and $\bar{W}_x(1,0)$ is given in the offered irreducible and faithful representation
\beq ||\bar{m}_x(j,k)||=1, ||\bar{W}(1,0)||=1, ||\bar{W}(0,1)||=1 \eeq

\section{Continuous extension of the shift}

In order to construct the continuous extension of the shift we concentrate on  the representation of the tracial state $\omega(AB)=\omega (BA),$ where the automorphisms used in the crossed product are unitarily implemented. Especially the gauge automorphism preserving $W_x(1,0) \quad \forall x\in Z$ is unitarily implemented by $U(1,0)$ with $U(1,0)^d =1$, though this operator does not belong to the quasilocal algebra but leaves the vector implementing the tracial state invariant. Therefore the Hilbertspace  $H$ can be decomposed into the orthogonal sum of $H_j ,j=0,  j-1$. With $$W_0(1,0)^j=\bar{W}_0(j,0),$$ we can with appropriate linear combinations create the isometries corresponding to matrix units $m_0(j,k)$ and their space translations that act as
\beq m_0(j,k)H_l \in \delta _{kl} H_j \eeq
 and similar for $\bar{m}_x(k,l).$

The subspaces $H_j$ are translation invariant. The gauge automorphism acts on them as multiplication with spectrum $[0,1).$ It can be combined with the reflection and the combined automorphism coincides with the continuous extension of $\beta _+\beta _-$ and is again unitarily implementable.
According to (11) $\bar{m}_x(k,j)$ as well as $m_x(r,s)$ act as isometries between the subspaces  $H_j$ to $H_k$  which due to the phase factor depend on the location of the matrix units.
In order to extend the discrete shift to a continuous automorphism group we are inspired by the relation  $[x,p] =i$ in one particle space. It can be combined with a rotation by defining \beq x=[x]+(x), \quad, [x]\in Z, \quad (x)\in [0,1) .\eeq
Then \beq e^{ip \delta }f([x],(x)) e^{-ip\delta }=f([x+\delta], (x+\delta ))\eeq
with the boundary condition $f(n,1)=f(n+1,0).$
Further we can define
\beq \gamma (\alpha )\bar{W}_x (1,0) =\bar{W}_x(\alpha )=\bar{W}_x(cos\alpha, sin\alpha), \quad x\in Z\eeq
This enables us to extend the relation between shift and rotation in the one particle space to our algebra by defining
\beq \bar{W}(x)= \bar{W}_{[x]}(cos(x),sin (x)),\quad x=[x]+(x), [x] \in Z, (x) \in [0,1)\eeq
Considering the shift to be implemented by $e^{iPn}$ in the Hilbertspace of the representation and $\gamma (\alpha )$ by $e^{iG\alpha}$ we can combine them to
\beq e^{i\delta \bar{P}}=e^{i[\delta ]P}e^{i((\delta )G}, \quad \delta =[\delta ]+(\delta ), [\delta ] \in Z, (\delta )\in (0,1)\eeq
This operator is well defined if it satisfies the desired boundary conditions. Here the differences between the two extensions of the quasilocal gauge invariant algebras appear. At the points $\delta =j/d \quad j=1,,d$ we move from $H_k$ to $H_{j+k}$. For the lattice algebra for $j=d$ the operators commute and the boundary condition is not satisfied, whereas for the Fermi algebra they anticommute according to (9). This makes it possible to transfer the boundary condition to the demand that $f(n,(x))$ is an odd function in $n$.

We can interpret this observation with the construction of the Fermi algebra. The automorphism that was combined with $W_x(0,1)$ is antisymmetric with respect to reflection at the point $x$. If we want to extend the discrete shift to the continuous shift offered in (16)again we have to do it by rotation at the points $y\neq x$ but in order that the new automorphism remains quasilocal the rotation must act in different direction under reflection.

For $d=2$ we are accustomed to write $\sum _x f(x)\bar{m}_x(1,2)=a(f)$ with $f\in l^2(x)$ and the identification $l^2(x)=L_{[0,1]}(p).$ Now again we can describe the matrix units to act on $L^2_{[0,1]}(p)$ as \beq \bar{m}_(j,k) f(p)= f(p+\frac{j-k}{d}).\eeq

\section{Construction of a time evolution}
  A time evolution can be constructed with local hamiltonians as discussed in \cite{BR} by using perturbation. It acts on $\M$ and under the assumptions that the local hamiltonians are gauge invariant it reduces to a time evolution on $\A_.$ If it satisfies the relevant commutation relations similar to those for the spacial translation, namely $\tau _t \gamma _{\beta } \tau _{-t} \gamma _{\beta }^{-1}$, then it can be extended to a time evolution for $\M_{\beta }$ . This works if the time evolution is constructed by a gauge invariant local hamiltonian. Similarily we can start with a time evolution on $\M_{\beta }$ under which  $\A $ is stable and satifies that $\tau _t \gamma \tau _{-t}\gamma ^{-1}$ remains an inner automorphism. In \cite{AMa} this  correspondence could be used to characterize its behaviour on $\M$ inherited from its transparent behaviour on $\M_{\beta }$ where it acted as a quasifree evolution.

  According to (17) quasifree evolutions can be defined also for general d-grading with keeping the action of the matrix elements by defining \beq\tau _t\bar{m}(j,k)f(p)=\bar{m}(j,k)f(e^{ih(p)t}p)\eeq
  with $h(p)$ a function in the one-particle space. Asymptotic behavior of the time evolution is given if the time evolution in the one particel space has absolutely continuous spectrum.
  This can be generalized to the total algebra $\M_{\beta}$ if we now consider $\M _{\beta }$ to be  defined as being generated by the products $\bar{m}(j,k)$ using their commutation relations. That these two definitions coincide is guaranted by the crossed product construction. But this is not sufficient for reducing it to a time evolution on $\A.$ As shown in \cite{MN} the continuous extension of the shift does not define an automorphism on $\A$ and $\M.$ This only holds with restrictions on $h(p).$
  In fact $\M_{\beta}$ being defined as crossed product of $\A$ an automorphism on $\M_{\beta }$ has to correspond to a variation how it respects the quasilocality of $\A.$ $\gamma _{\beta }^{-1}\tau _t \gamma {\beta }$ must reduce to an automorphism on $\A$ though not an inner one. This leads to conditions on $h(p).$ An automophism given by a local hamiltonian quadratic in the matrix units is given by
\beq \frac{d}{dt} \tau _t A=i\sum _z \sum _x h(x)[\bar{W}_z(0,1)\bar{W} _{x+z}(0,-1)+ c.c, A]\eeq
The commutator is given by a local operator and therefore on the basis of $\M$  perturbation theory gives a well defined time evolution that by gauge invariance can be restricted to a time evolution on $\A$.

It becomes more transparent if we evaluate it by its action on $\bar{W}(0,1)$ and afterwards transfer it to $\A.$
\beq [ \bar{W}_0(0,-1)\bar{W}_x(0,1), \bar{W}_z(0,1)]= e^{2\pi i \frac{j_++j_-}{d}}[W_z(1,0)^{j_++j_-},\bar{W}_z(0,1)], \quad 0<z<x \eeq
$$ [ \bar{W}_0(0,-1)\bar{W}_x(0,1), \bar{W}_0(0,1)]= cos(2\pi \frac{j_+}{d})W_x(1,0)^{j_-}\bar{W}_x (0,1)$$
$$ [ \bar{W}_0(0,-1)\bar{W}_x(0,1)+c.c, \bar{W}_x(0,1)]= cos(-2\pi \frac{j_+}{d})W_x(1,0)^{-j_-}\bar{W}_0 (0,1)$$
If we choose $j_+=j_-$ then this reduces to
\beq \frac{d}{dt} \tau _t|_0 \bar{W}(0,1)= cos (2\pi j_+)(\bar{W}_x(2j_+,1)+\bar{W}_{-x}(2j_+,1)) \eeq
This enables us to use the expression of $\bar{W}$ as operators smeared over a linear space and accordingly express the time evolution in this linear space.

\begin{definition} $\bar{W}(f(x,j))= \sum _{x,j}f(x,j) \bar{W}_x(j,1)$ \end{definition}

Summing over interactions given by $$\sum h(x-y )\bar{W}_x(0,1)\bar{W}_y(0,-1)$$
we can express the time evolution by
$$\tau_t \bar{W}(f)=\bar{W}(f_t)$$
where \beq\frac{d}{dt}f_t(x,j) = \sum_y \sum _k\int d p \int d\alpha e^{ip(x-y)}e^{2\pi i \alpha (j-k)}\int dq e^{iqz}(h(y-z)f(z,k)\eeq
With the necessary demand on locality of $h(x)$ the time translated $f(x)$ after Fourier transition satisfies the continuity properties of (17).
Finitely \beq  \tau _t W(1,0) =\tau _te^{2\pi i/d} \bar{W}(0,1) \bar{W}(1,-1)\eeq
and is given by the time evolution of $\bar{W}(j,1)$ together with the fact that the time evolution in the above combination preserves locality so that the right sight of the equation is well defined.

So far we have constructed a special class of time evolutions in the sense of \cite{BR}. But for this class of time evolutions we can control the asymptotic behaviour. Evidently the automorphism implementing the time evolution in the tracial state has, apart from the vector implementing the tracial state,  absolutely continuous spectrum as can be observed from (22). Matrix units do not commute asymptotically neither for the shift nor for the time evolution. But in the combination of the matrix units presented in (17) the commutator is given by the convolution of (22) which in Fourier space reduces to a product with a hamiltonian with absolutely continuous spectrum and therefore decreases asymptotically to $0.$ As for 2-grading the gauge invariant algebra with the time evolution given in (19) inherits this asymptotic behaviour.

In addition we can construct KMS-states corresponding to the quasifree evolution corresponding to $h(p)$ using the results of \cite{BR}. But since time evolutions corresponding to different $h(p)$ commute with one another it follows that all these states are invariant under all quasifree evolutions that are space translation invariant. These states can be considered to be the analog of quasifree states for d-grading. Again they are fixed by a two-point function and are given by the sum of products of these two-point functions with phase factors determined by the commutation relations. As in the tracial state the two-point function corresponding to $m_x(j,k)\tau _tm_y(s,l) $ decreases to $0$ due to the absolutely continuous spectrum of $h(p).$

In \cite{N1} it was stated, that interacting time evolutions on the spin algebra cannot be norm asymptotically abelian. But this was proven by showing that it does not hold for the very special operators $m_x(j,k)$. As for the XY-model of \cite{AMa} it holds on the gauge invariant subalgebra. This is here generalized to the gauge invariant subalgebra for d-grading. The results of \cite{N1} imply that as for the XY-model the total algebra is not asymptotically abelian.

\section{d-grading in relation to kd-grading}
Constructing $\M$ we can also start with the full matrix algebra $M_o\otimes... M_k$ which has dimension $kd$ and take the tensorproduct with its translates $\sigma ^k$. Therefore we can construct $\M$ also as crossed product of $\A (k)$ as the algebra invariant under a gauge group of dimension $kd$ with the corresponding gauge automorphism satisfying $\gamma ^{kd}=1.$ Here $\A(k)\subset \M$ being invariant under a larger gauge group is a subalgebra of $\A\subset \M.$

The crossed product extensions give both $\M$ and therefore coincide. On $\A$ the shift $\sigma $ is defined and can be extended to $\M.$ On $\A(k)$ only $\sigma ^k$ is defined and its extension how it acts on $W_{kd,kx}(0,1)$ is given for $\sigma ^k.$ However as element of $\M$ also  $\sigma W_{kd,kx}(0,1) $ is given. Based on the crossed product extension it can be constructed in exactly the same way as the continuous extension of the shift for the Fermi-algebra $\M_F$, now not by a continuous rotation but by a discrete rotation with corresponding boundary conditions.

This observation finds its analog for the Fermi-type-extensions. We start with $\A (kd)$ and extend it to $\M _{\beta}.$ This corresponds to a refinement of the steps in (17) corresponding to the increased number of matrix units. However the action of the shift on $f(p)$
referring just to the boundary condition is not effected by $k$. Again it follows that $\M_{\beta }$ does not depend on the choice of the size of the steps.

The main motivation to study the Fermitype-extension was the possibility to control asymptotic time behaviour on $\M$ for a larger class of time evolutions. Quasifree evolutions act as automorphisms on $\M_F$ though defined by referring to the matrix units remain unchanged under a coarse-graining of the matrix units. They are norm asymptotically  abelian on $\A(d)$ respectively on $\A(kd)$ if they can be reduced to automorphisms on the subalgebra. This is a stronger demand for $\A(d)$. However already for $\A(kd)$ we can increase the demand by the natural restriction that the time evolution has to commute with $\sigma $ and not only with $\sigma ^k$. Characterizing the time evolution by $e^{ih(p)}$
and comparing it with the calculation in (22) the restrictions for $h(p)$ at $p=l/k$ coincide so that the passage to higher grading does not offer new examples of time evolutions that are asymptotically abelian on the subalgebra $\A(d).$

\section{Extension to higher dimensional lattices}
For one dimensional lattices the connection of the spin algebra and the Fermi algebra is a well established fact. Both algebras can be easily generalized to algebras on higher dimensional lattices. However the mapping between the subalgebras $\A$ that again are defined by gauge invariance is given by the construction using $\beta$ and we have to search for an analog in higher dimensions. Again the fact that for the Fermi-algebra the shift can be continuously extended offers a possible connection.

For the one dimensional lattice it was essential that the automorphism $\beta $ turned into $\beta ^{-1}$ under reflection so that shifting one of them by a finite amount their combination acted trivially outside of a finite region.  For the continuous extension
of the shift we added rotation with appropriate boundary condition that also changed its sign under reflection. The Fermi algebra also in higher dimension offers a quasilocal even subalgebra respectively a quasilocal subalgebra under a d-gauge group. More precisely the algebra built by creation- and annihilation operators respectively the matrix units $\bar{m}_x(j,k)$ located in a finite region of the lattice is isomorphic to the full matrix algebra of size determined by the number of points in the region. On the algebra we have both space translation and rotation, nontrivially connected. For both automorphisms we define reflection, for spacetranslation with respect to a fixed point, for rotation with respect to the d-dimensional matrix-algebra at a point. Together they are a replacement of  the explicit $\beta $ on the one dimensional lattice. Combining reflection and space translated reflection they reduce to a mapping on local algebras.

\section{Conclusion}
Based on the fact that the spin-algebra and the Fermi-algebra on the one dimensional lattice are strongly related and this relation is a useful tool for further investigations we generalized the construction to higher dimensional gauge groups as well as to higher dimensional lattices. The commutation relations become slightly more difficult, but it turned out, that the relevant basis of the relation between spin-algebra and Fermi-algebra is the reflection operation, acting both on the lattice as well as on the local algebra. On the Fermi-algebra quasifree evolutions can be defined, either as evolution in an $l^2$ space or by the explicit operator that generates the evolution in the tracial state. Both algebras contain a common gauge invariant subalgebra, though with a mapping that is not an automorphism but is defined by the different automorphisms that create over the crossed product construction the total algebras . Automorphisms on the total algebras can be transferred from one algebra to the other it they preserve the common gauge invariant subalgebra but only if they satisfy appropriate conditions with respect to the mapping.

\bibliographystyle{plain}

\begin{thebibliography}{99}
\bibitem{A} H. Araki: Publ.RIMS, Kyoto Univ {\bf20} 277 (1984)
\bibitem{AMo} H. Araki, H. Moriya: Rev. Math. Phys. {\bf15} 93 (2003)
\bibitem{AMa} H. Araki, T. Matsui: Commun. Math. Phys. {\bf101} 213 (1985)
\bibitem{O} Y. Ogata: Commun. Math. Phys. {\bf245/3} 577 (2003)
\bibitem{RMO} J.G. Rubio, A. Molnar, Y. Ogata: Classifying  symmetric and symmetry broken spin chain phases with anomalous gropup action arXiv:2403.18573
\bibitem{N1} H. Narnhofer: Asymptotics for automorphisms of quantum systems on the lattice, arXiv:2405.19696
\bibitem{N2} H. Narnhofer: Interacting Fermi systems in the tracial state, arXiv:2406.04977
\bibitem{BR} O. Bratteli, D.W. Robinson: Operator Algebras and Quantum Statistical Mechanics I and II, Springer (1981)

\bibitem{MN} H. Moriya, H. Narnhofer: Continuous extension of the discrete shift translations on one-dimensional quantum lattice systems,
 arXiv 2405.01001 math-ph


\end{thebibliography}

\end{document}